 \documentclass[journal]{IEEEtranTCOM}
\usepackage{amssymb}
\usepackage{multicol}
\usepackage{subfloat}

\normalsize

\ifCLASSINFOpdf
\else
\fi

\begin{document}
\newtheorem{theorem}{Theorem}
\newtheorem{proposition}{Proposition}
\newtheorem{lemma}{Lemma}
\newtheorem{definition}{Definition}
\newtheorem{corollary}{Corollary}
\newtheorem{remark}{Remark}
\newtheorem{example}{Example}
%
\title{The Weight Distributions of a Class of Cyclic Codes with Three Nonzeros over $\mathbb{F}_3$\thanks{
This work is supported in part by the National Key Basic Research and Development Plan of China under Grant 2012CB316100, and
the National Natural Science Foundation of China under Grants 61271222, 60972033.}}

\author{ Xiaogang Liu and Yuan Luo
\thanks{X. Liu and Y. Luo are with the
Computer Science and Engineering Department,
Shanghai Jiao Tong University, Shanghai 200240, China e-mail: liuxg0201@163.com and yuanluo@sjtu.edu.cn.}}

\markboth{IEEE Transactions on Communications}%
{Submitted paper}

\maketitle

\begin{abstract}
Cyclic codes have efficient encoding and decoding algorithms. The decoding error probability and the undetected error probability are usually bounded by or given from the weight distributions of the codes. Most researches are about the determination of the weight distributions of cyclic codes with few nonzeros, by using quadratic form and exponential sum but limited to low moments. In this paper, we focus on the application of higher moments of the exponential sum to determine the weight distributions of a class of ternary cyclic codes with three nonzeros, combining with not only quadratic form but also MacWilliams' identities. Another application of this  paper is to emphasize the computer algebra system Magma for the investigation of the higher moments. In the end, the result is verified by one example using Matlab.
\end{abstract}


\IEEEpeerreviewmaketitle

\section{Introduction}\label{secI}

 Cyclic codes have a lot of applications in communication system, storage system and computers. The decoding error probability and the undetected error probability are closely related with the weight distributions. For example, permutation decoding,  majority decoding, locator decoding, decoding from the covering polynomials and so on \cite{K001,PH001,SF001,SFT001}. In general the weight distributions are complicated \cite{MS01} and difficult to be determined.
In fact, as shown in \cite{MR001} and \cite{V01}, the problem of computing weight distribution of a cyclic code is connected to the evaluation of certain exponential sums, which are generally hard to be determined explicitly.
For more researches, refer to \cite{D001,DY001,FY001,M001} for the irreducible case, \cite{DL001,FL001,LTW001,V001} for the reducible case, and \cite{LHF001,LL002,X002,ZDJZ001} for recent studies. Especially, for related problems in the binary case with two nonzeros, refer to \cite{JH001,JHK001} and \cite{BM001}.

  In this paper, we focus on the application of higher moments of the exponential sum to determine the weight distribution of a class of ternary cyclic codes with three nonzeros, combining with not only quadratic form but also MacWilliams' identities, with the help of the computer algebra system Magma. 

Let $p$ be a prime. A linear $[n,k,d;p]$ code is a $k$-dimensional subspace of $\mathbb{F}_p^n$ with minimum (Hamming) distance $d$. An $[n,k]$ linear code $\mathcal{C}$ over $\mathbb{F}_p$ is called cyclic if $(c_0,c_1,\cdots,c_{n-1})\in \mathcal{C}$ implies that $(c_{n-1},c_0,c_1,\cdots,c_{n-2})\in \mathcal{C}$ where $\mbox{gcd}(n,p)=1$.
 By identifying the vector $(a_0,a_1,\ldots,a_{n-1})\in \mathbb{F}_p^m$ with
 \[
a_0+a_1x+\cdots a_{n-1}x^{n-1}\in \mathbb{F}_p[x]/(x^n-1),
\]
any linear code $\mathcal{C}$ of length $n$ over $\mathbb{F}_p$  represents a subset of $\mathbb{F}_p[x]/(x^n-1)$ which is a principle ideal domain. The fact that the code is cyclic is equivalent to that the subset is an ideal. The unique monic polynomial $g(x)$ of minimum degree in this subset is the generating polynomial of $\mathcal{C}$, and it is a factor of $x^n-1$. When the ideal does not contain any smaller nonzero ideal, the corresponding cyclic code $\mathcal{C}$ is called a minimal or an irreducible code.
    For any $v=(c_0,c_1,\cdots,c_{n-1})\in \mathcal{C}$, the weight of $v$ is $wt(v)=\#\{c_i\not=0,i=0,1,\ldots,n-1\}$.

 The weight enumerator of a code $\mathcal{C}$ is defined by
\[
1+A_1x+A_2x^2+\cdots+A_nx^n,
\]
where $A_i$ denotes the number of codewords with Hamming weight $i$.
The sequence $1,A_1,\cdots,A_n$ is called the weight distribution of the code, which is an important parameter of a linear block code.

Assume that $p=3$ and $q=p^m$ for an even integer $m$.
 Let $\pi$ be a primitive element of $\mathbb{F}_q$. In this paper, Section \ref{SecII} presents the basic notations and preliminaries about cyclic codes. Section \ref{SecIII} determines the weight distributions of a class of cyclic codes over $\mathbb{F}_3$ with nonzeros $\pi^{-2},\pi^{-4},\pi^{-10}$, and they are verified by using Matlab. Note that the length of the cyclic code is $l=q-1=3^m-1$. Final conclusion is in Section \ref{SecIV}. This paper is the counterpart of our another result in \cite{LL002}.

\section{Preliminaries\label{SecII}}

In this section, relevant knowledge from finite fields \cite{LH001} is presented for our study of cyclic codes. It is about the calculations of exponential sums, the sizes of cyclotomic cosets and the ranks of certain quadratic forms.
First, some known properties about the codeword weight are listed. Then Lemma \ref{FL002}, Lemma \ref{ES02} and Lemma \ref{R001} are about the calculations of exponential sums. Finally,  Lemma \ref{GCD002}, Lemma \ref{RQ002} and Corollary \ref{RQ02} are about the ranks of relevant quadratic forms.

 Let $p$ be an odd prime, $m$ be a positive integer and $\pi$ is a primitive element of $\mathbb{F}_q$. Assume the cyclic code $\mathcal{C}$ over $\mathbb{F}_p$ has length $l=q-1=p^m-1$ and non-conjugate nonzeros $\pi^{-s_\lambda}$, where $1\leq s_\lambda \leq q-2 (1\leq \lambda \leq \iota)$.  Then the codewords in $\mathcal{C}$ can be expressed by
\begin{equation}\label{CW001}
c(\alpha_1,\ldots,\alpha_{\iota}) = (c_0,c_1,\ldots,c_{l-1}) \ \ (\alpha_1,\ldots,\alpha_{\iota}\in \mathbb{F}_q),
\end{equation}
where $c_i = \sum\limits_{\lambda =1}^{\iota}\mbox{Tr}(\alpha_{\lambda}\pi^{is_{\lambda}}) (0\leq i\leq l-1)$ and $\mbox{Tr}:\mathbb{F}_q\rightarrow \mathbb{F}_p$ is the trace mapping from $\mathbb{F}_q$ to $\mathbb{F}_p$. Therefore the Hamming weight of the codeword $c=c(\alpha_1,\ldots,\alpha_{\iota})$ is:

\begin{equation}\label{C01}
\begin{array}{ll}
w_H(c)&=\# \{i|0\leq i \leq l-1, c_i\not= 0\}\\
      &=l-{l\over p}-{1\over p}\sum\limits_{a=1}^{p-1}\sum\limits_{x\in \mathbb{F}_q^*}{\zeta}_p^{\mbox{Tr}(af(x))}\\
      &=p^{m-1}(p-1)-{1\over p}\sum\limits_{a=1}^{p-1}S(a\alpha_1,\ldots,a\alpha_{\iota})\\
      &=p^{m-1}(p-1)-{1\over p}R(\alpha_1,\ldots,\alpha_{\iota})
\end{array}
\end{equation}
where  ${\zeta_p}=e^{{2\pi i}\over p}$ ($i$ is imaginary unit), $f(x)=\alpha_1x^{s_1}+\alpha_2x^{s_2}+\cdots +\alpha_{\iota}x^{s_{\iota}}\in \mathbb{F}_q[x], \mathbb{F}_q^*=\mathbb{F}_q\backslash \{0\}$,
\begin{equation}\label{ES00001}
S(\alpha_1,\ldots,\alpha_{\iota}) = \sum\limits_{x\in \mathbb{F}_q}{\zeta}_p^{\mbox{Tr}\left(\alpha_1x^{s_1}+\cdots +\alpha_{\iota}x^{s_{\iota}}\right)},
\end{equation}
and $R(\alpha_1,\ldots,\alpha_{\iota})=\sum\limits_{a=1}^{p-1}S(a\alpha_1,\ldots,a\alpha_{\iota})$.

For general functions of the form $
f_{\alpha,\ldots,\gamma}(x) = \alpha x^{p^i+1} + \cdots + \gamma x^{p^j+1}$
 where $0\leq i,\ldots, j \leq  \lfloor{{m}\over 2}\rfloor$, there are quadratic forms
 \begin{equation}\label{QF01}
 F_{\alpha,\ldots,\gamma}(X)
 \end{equation}
  and corresponding symmetric matrices
  \begin{equation}\label{QF02}
  H_{\alpha,\ldots,\gamma}
  \end{equation}
  satisfying that $F_{\alpha,\ldots,\gamma}(X)=XH_{\alpha,\ldots,\gamma}X^T=\mbox{Tr}(f_{\alpha,\ldots,\gamma}(x)).$

\begin{lemma}(Lemma 1, \cite{FL001})\label{FL002}
\begin{enumerate}
\renewcommand{\labelenumi}{$($\mbox{\roman{enumi}}$)$}
\item
For the quadratic form $F(X) = XHX^T$, 
\[
\sum\limits_{X\in \mathbb{F}_p^m}\zeta_p^{F(X)} =
\left\{
\begin{array}{ll}
\left({\Delta \over p}\right)p^{m-r/2} & \mbox{if} \ \ p\equiv 1 \ (\mbox{mod} \ 4),\\
i^r\left({\Delta \over p}\right)p^{m-r/2} & \mbox{if} \ \ p \equiv 3 \ (\mbox{mod} \ 4).
\end{array}
\right.
\]
\item
For $A = (a_1,\ldots,a_m)\in \mathbb{F}_p^m$, if \ $2YH+A=0$ has solution $Y=B \in \mathbb{F}_p^m$, then
\begin{equation}
\sum_{X\in \mathbb{F}_p^m}{\zeta}_p^{F(X)+AX^T} = {\zeta}_{p}^c\sum_{X\in \mathbb{F}_p^m}{\zeta}_p^{F(X)}
\end{equation}
where $c={1\over 2}AB^T\in \mathbb{F}_p$.
Otherwise $\sum\limits_{X\in \mathbb{F}_p^m}{\zeta}_p^{F(X)+AX^T} = 0$.
\end{enumerate}
Here
 \[
 \left({\Delta \over p}\right)
\]
 denotes the Legendre symbol.
\end{lemma}

 Lemma \ref{ES02} is from \cite{LL001}, Lemma \ref{R001} is from \cite{LL002}, also refer to \cite{FL001} for the calculations of exponential sums that will be needed in the sequel.
\begin{lemma}\label{ES02}
For the quadratic form $F_{\alpha,\ldots,\gamma}(X) = XH_{\alpha,\ldots,\gamma}X^T$ corresponding to $f_{\alpha,\ldots,\gamma}(x)$, see (\ref{QF01})
\begin{enumerate}
\renewcommand{\labelenumi}{$($\mbox{\roman{enumi}}$)$}
\item
if the rank $r_{\alpha,\ldots,\gamma}$ of the symmetric matrix $H_{\alpha,\ldots,\gamma}$ is even, which means that $S(\alpha,\ldots,\gamma) = \varepsilon p^{m-{r_{\alpha,\ldots,\gamma}\over 2}}$, then
\[
R(\alpha,\ldots,\gamma) =\varepsilon (p-1) p^{m-{r_{\alpha,\ldots,\gamma}\over 2}}; 
\]
\item
if the rank $r_{\alpha,\ldots,\gamma}$ of the symmetric matrix $H_{\alpha,\ldots,\gamma}$ is odd, which means that $S(\alpha,\ldots,\gamma) = \varepsilon \sqrt{p^*}p^{m-{{r_{\alpha,\ldots,\gamma}+1}\over 2}}$, then
\[
R(\alpha,\ldots,\gamma) = 0
\]
\end{enumerate}
where $ \varepsilon = \pm 1$ and $p^*=\left({{-1} \over p}\right)p$.
\end{lemma}

\begin{lemma}\label{R001}
 Let $F_{\alpha,\ldots,\gamma}(X) = XH_{\alpha,\ldots,\gamma}X^T$ be the quadratic form corresponding to $f_{\alpha,\ldots,\gamma}(x)$, see (\ref{QF01}).
If the rank $r_{\alpha,\ldots,\gamma}$ of the symmetric matrix $H_{\alpha,\ldots,\gamma}$ is odd, then the number of quadratic forms with exponential sum $\sqrt{p^*}p^{m-{{r_{\alpha,\ldots,\gamma}+1}\over 2}}$ equals the number of quadratic forms with exponential sum $-\sqrt{p^*}p^{m-{{r_{\alpha,\ldots,\gamma}+1}\over 2}}$ where $p^*=\left({{-1} \over p}\right)p$.
\end{lemma}

The cyclotomic coset containing $s$ is defined to be
\begin{equation}\label{CC01}
\mathcal{D}_s=\{s,sp,sp^2,\ldots,sp^{m_s-1}\}
\end{equation}
where $m_s$ is the smallest positive integer such that $p^{m_s}\cdot s \equiv s \ (\mbox{mod} \ p^m-1)$. In the following, Lemma \ref{GCD002} and Lemma \ref{RQ002} are from \cite{LL001}, also refer to \cite{CH001} for the binary case of Lemma \ref{GCD002}.
\begin{lemma}\label{GCD002}
If $m=2t+1$ is odd, then for $l_i=1+p^i$, the cyclotomic coset $\mathcal{D}_{l_i}$ has size
\[
|\mathcal{D}_{l_i}|=m, \ \ \ 0\leq i \leq t.
\]
If $m=2t+2$ is even, then for $l_i=1+p^i$, the cyclotomic coset $\mathcal{D}_{l_i}$ has size
\[
|\mathcal{D}_{l_i}| =
\left\{
\begin{array}{ll}
m, & 0\leq i \leq t \\
m/2, & i=t+1.
\end{array}
\right.
\]
 \end{lemma}

For $f_d'(x) =\alpha_0 x^2+\alpha_1x^{p+1}+\cdots+\alpha_dx^{p^d+1}$ with corresponding quadratic form $F_d'(X)=\mbox{Tr}(f_d'(x))=XH_d'X^T$ where $(\alpha_0,\alpha_1,\ldots,\alpha_d)\in \mathbb{F}_q^{d+1}\backslash\{(0,0,\ldots,0)\}$, the following result is about its rank.
\begin{lemma}\label{RQ002}
Let $m$ be a positive integer, $0\leq d\leq \lfloor{m\over 2}\rfloor$. The rank $r_d'$ of the symmetric matrix $H_d'$ satisfies $r_d'\geq m-2d$.
\end{lemma}

The following corollary is a special case of Lemma \ref{RQ002}.
\begin{corollary}\label{RQ02}
The rank $r_2'$ of the symmetric matrix $H_2'$ corresponding to $f_2'(x) =\alpha_0 x^2+\alpha_1x^{p+1}+\alpha_2x^{p^2+1}$ has five possible values:
\[
m,m-1,m-2,m-3,m-4.
\]
\end{corollary}

 {\bf Note that in Section \ref{SecIII}, a nonzero solution of a equation system means that all the variable values are nonzero.}

\section{Main Results}\label{SecIII}

In this section, the main results of this paper are obtained, that is the weight distribution of the cyclic code $C$ with nonzeros $\pi^{2}, \pi^{p+1}$ and $\pi^{p^2+1}$ for the case when $m$ is even, here $p=3$.
For this, the first five moments of exponential sum $S(\alpha,\beta,\gamma)$ are computed in Subsections \ref{Sec3.1}, \ref{Sec3.2}, \ref{Sec3.3}, and the MacWlliams' identities are calculated in Subsection \ref{Sec3.4}.

\subsection{The First Three Moments of $S(\alpha,\beta,\gamma)$}\label{Sec3.1}


For an odd prime $p$ and even integer $m$, this subsection calculates the first three moments of the exponential sum $S(\alpha,\beta,\gamma)$ (equation (\ref{ES00001})), see Lemma \ref{ESL001} and its another form Lemma \ref{ES002}, where the analysis of the third moment bases on the property of Lemma \ref{ESL002}.

\begin{lemma}(Theorem 6.26., \cite{LH001}) \label{SE01}
Let $f$ be a nondegenerate quadratic form over $\mathbb{F}_q$, $q$ odd, in an even number $n$ of indeterminates. Then for $b\in \mathbb{F}_q$ the number of solutions of the equation $f(x_1,\ldots,x_n)=b$ in $\mathbb{F}_q^n$ is
\[
q^{n-1}+\upsilon (b)q^{(n-2)/ 2} \eta \left((-1)^{n/ 2}\Delta\right)
\]
where $\upsilon (b)=-1$ for $b\in \mathbb{F}_q^*$, $\upsilon (0)=q-1$, $\eta$ is the quadratic character of $\mathbb{F}_q$ and $\Delta =\mbox{det}(f)$.
\end{lemma}

\begin{lemma}\label{ESL002}
Let $p$ be an odd prime, $q=p^m$ and $a\in \mathbb{F}_{q}^*$. Then the solutions of the following equation in $\mathbb{F}_q^2$
\begin{equation}\label{SE001}
x^2+y^2=a
\end{equation}
have the form
\begin{equation}\label{SE002}
x={1\over 2}s\left(\theta+\theta^{-1}\right), y={1\over 2}st\left(\theta-\theta^{-1}\right)
\end{equation}
where $s, t, \theta \in \mathbb{F}_{q^2}^*$ and $s^2=a, t^2=-1$.
\end{lemma}

\begin{IEEEproof}
First, it can be checked that the pairs $x, y$ given by the lemma satisfy equation (\ref{SE001}). Second, according to Lemma \ref{SE01}, the number of solutions of equation (\ref{SE001}) in $\mathbb{F}_{q^2}^2$ is $q^2- \eta \left(-1\right)=q^2-1$ since $-1$ is a quadratic residue of $\mathbb{F}_{q^2}^*$. Furthermore, when $\theta$ varies through the nonzero elements of $\mathbb{F}_{q^2}$, $(x,y)$ in (\ref{SE002}) gives all the solutions of (\ref{SE001}) in $\mathbb{F}_{q^2}^2$ including those solutions in the subfield $\mathbb{F}_{q}^2$. In fact, $\theta_1+\theta_1^{-1}=\theta_2+\theta_2^{-1}$ and $\theta_1-\theta_1^{-1}=\theta_2-\theta_2^{-1}$ imply $\theta_1=\theta_2$. So for $\theta_1\not=\theta_2$, $(x_1,y_1)\not=(x_2,y_2)$.
\end{IEEEproof}

\begin{lemma}\label{ESL001}
Let $p$ be an odd prime satisfying $p\equiv 3 \ \mbox{mod}\ 4$ and $q=p^m$. 
 Then there are the following results about the exponential sum $S(\alpha,\beta,\gamma)$ (equation (\ref{ES00001})) corresponding to $f_2'(x) =\alpha x^2+\beta x^{p+1}+\gamma x^{p^2+1}$ 
\begin{enumerate}
\renewcommand{\labelenumi}{$($\mbox{\roman{enumi}}$)$}
\item
$\sum\limits_{\alpha,\beta,\gamma\in \mathbb{F}_q} S(\alpha,\beta,\gamma)=p^{3m}$
\item
$\sum\limits_{\alpha,\beta,\gamma\in \mathbb{F}_q} S(\alpha,\beta,\gamma)^2=p^{3m}$
\item
$\sum\limits_{\alpha,\beta,\gamma\in \mathbb{F}_q} S(\alpha,\beta,\gamma)^3=\left((p+1)(p^m-1)+1\right)p^{3m}$.
\end{enumerate}
\end{lemma}

\begin{IEEEproof}
From definition, changing the order of summations,
(i) can be calculated as follows
\[
\begin{array}{lll}
&&\sum\limits_{\alpha,\beta,\gamma\in \mathbb{F}_q} S(\alpha,\beta,\gamma)\\
&=&\sum\limits_{\alpha,\beta,\gamma\in \mathbb{F}_q}\sum\limits_{x\in\mathbb{F}_q}\zeta_p^{\mbox{Tr}\left(\alpha x^2+\beta x^{p+1}+\gamma x^{p^2+1}\right)} \\
 &=&\sum\limits_{x\in \mathbb{F}_q}\sum\limits_{\alpha \in \mathbb{F}_q}\zeta_p^{\mbox{Tr}\left(\alpha x^2\right)}\sum\limits_{\beta \in \mathbb{F}_q}\zeta_p^{\mbox{Tr}\left(\beta x^{p+1}\right)}\sum\limits_{\gamma \in \mathbb{F}_q}\zeta_p^{\mbox{Tr}\left(\gamma x^{p^2+1}\right)}   \\
 &=&\sum\limits_{\stackrel{\alpha \in \mathbb{F}_q}{x=0}}\zeta_p^{\mbox{Tr}\left(\alpha x^2\right)}\sum\limits_{\stackrel{\beta \in \mathbb{F}_q}{x=0}}\zeta_p^{\mbox{Tr}\left(\beta x^{p+1}\right)}\sum\limits_{\stackrel{\gamma \in \mathbb{F}_q}{x=0}}\zeta_p^{\mbox{Tr}\left(\gamma x^{p^2+1}\right)} \\
 &=&q^3=p^{3m}.
 \end{array}
\]

Equation (ii) can also be calculated in this way
\[
\begin{array}{lll}
&&\sum\limits_{\alpha,\beta,\gamma\in \mathbb{F}_q} S(\alpha,\beta,\gamma)^2 \\
&=&\sum\limits_{x,y\in \mathbb{F}_q}\sum\limits_{\alpha \in \mathbb{F}_q}\zeta_p^{\mbox{Tr}\left(\alpha\left(x^{2}+y^{2}\right)\right)}\sum\limits_{\beta \in \mathbb{F}_q}\zeta_p^{\mbox{Tr}\left(\beta\left(x^{p+1}+y^{p+1}\right)\right)}\\
&& \ \ \ \ \ \ \  \sum\limits_{\gamma \in \mathbb{F}_q}\zeta_p^{\mbox{Tr}\left(\gamma\left(x^{p^2+1}+y^{p^2+1}\right)\right)}\\
&=&M_2\cdot p^{3m}
\end{array}
\]
where $M_2$ is the number of solutions to the equation system
\begin{equation}\label{ES201}
\left\{
 \begin{array}{ll}
 x^2+y^2 &=0\\
 x^{p+1}+y^{p+1} &=0\\
 x^{p^2+1}+y^{p^2+1} &=0.
 \end{array}
\right.
\end{equation}
If $x,y\not=0$ satisfy the above system, then $({x\over y})^2=-1$ and $({x\over y})^{p+1}=-1$. Since $p\equiv 3 \ \mbox{mod}\ 4$, that is $p+1\equiv 0 \ \mbox{mod}\ 4$, we have $({x\over y})^{p+1}=(({x\over y})^2)^{{p+1}\over 2}=(-1)^{{p+1}\over 2}=1$, a contradiction. So, the only solution to above system is $x=y=0$ and $M_2=1$.


As to (iii), we have
\begin{equation}\label{CM02}
\sum\limits_{\alpha,\beta,\gamma\in \mathbb{F}_q} S(\alpha,\beta,\gamma)^3=M_3\cdot p^{3m}
\end{equation}
where
\begin{equation}\label{CM0001}
\begin{array}{lll}
M_3&=\#\{(x,y,z)\in \mathbb{F}_q^3 \ |& x^2+y^2+z^2=0,\\
          &       &x^{p+1}+y^{p+1}+z^{p+1}=0,\\
          &&x^{p^2+1}+y^{p^2+1}+z^{p^2+1}=0\}\\
 &=M_2+T_3\cdot (q-1),
\end{array}
\end{equation}
and $T_3$ is the number of solutions of
\begin{equation}\label{ES01}
\left\{
\begin{array}{ll}
x^2+y^2+1 &=0 \\
x^{p+1}+y^{p+1}+1 &=0 \\
x^{p^2+1}+y^{p^2+1}+1 &=0.
\end{array}
\right.
\end{equation}
To study equation system (\ref{ES01}), consider the last two equations. Canceling $y$ there is
\[
\left(x^{p+1}+1\right)^{p^2+1}=\left(x^{p^2+1}+1\right)^{p+1},
\]
after simplification, it becomes
\begin{equation}\label{ES0002}
\left(x^{p^2}-x^p\right)\left(x^{p^3}-x\right)=\left(x^{p}-x\right)^p\left(x^{p^3}-x\right)=0.
\end{equation}
From (\ref{ES0002}) it can be checked that $x\in \mathbb{F}_{p^3}$. In the same way, it implies that $y\in \mathbb{F}_{p^3}$. In this case,
$(x^{p^2+1}+y^{p^2+1}+1)^p=x^{p^3+p}+y^{p^3+p}+1=0 \Longleftrightarrow x^{p+1}+y^{p+1}+1=0$, so only the first two equations of (\ref{ES01}) are necessary to be considered.

 The following paragraph of proof is similar to what has been done in \cite{ZDJZ001}. For the first one of equation (\ref{ES01}), by Lemma \ref{ESL002} there exist $s, \theta \in \mathbb{F}_{p^6}$ such that $x={1\over 2}s(\theta+\theta^{-1})$ and $y={1\over 2}(\theta-\theta^{-1})$ where $s^2=-1$. Substituting to the second equation,
 \[
 \begin{array}{ll}
 {1\over 4}s^{p+1}\left(\theta^p+\theta^{-p}\right)\left(\theta+\theta^{-1}\right)+{1\over 4}\left(\theta^p-\theta^{-p}\right)\left(\theta-\theta^{-1}\right)+1=0.
 \end{array}
 \]
Since $p+1\equiv 0 \ \mbox{mod}\ 4$, $s^{p+1}=(-1)^{{p+1}\over 2}=1$. After simplification, the above equation becomes ${1\over 2}\left(\theta^{p+1}+\theta^{-(p+1)}\right)+1=0$. Set $\tau=\theta^{p+1}$, we have $\tau^2+2\tau+1=(\tau+1)^2=0$, that is
\begin{equation}\label{ES000201}
\tau=\theta^{p+1}=-1.
\end{equation}

Now, since $x\in \mathbb{F}_{p^3}$,
\[
\left({1\over 2}s\left(\theta+\theta^{-1}\right)\right)^{p^3}={1\over 2}s^{p^3}\left(\theta^{p^3}+\theta^{-p^3}\right)={1\over 2}s\left(\theta+\theta^{-1}\right).\]
 That is $\left(\theta^{p^3-1}+1\right)\left(\theta^{p^3+1}+1\right)=0$
 where $p^3-1 \equiv 2\ \mbox{mod}\ 4$. Also, because $y\in \mathbb{F}_{p^3}$, we can obtain $\left(\theta^{p^3-1}-1\right)\left(\theta^{p^3+1}+1\right)=0$. Combining the two results, we find that
\begin{equation}\label{ES000202}
\theta^{p^3+1}=-1.
\end{equation}

If $\theta_1$ and $\theta_2$ satisfy (\ref{ES000201}) and  (\ref{ES000202}), we have $({\theta_1\over \theta_2})^{p+1}=({\theta_1\over \theta_2})^{p^3+1}=1$. Since $\mbox{gcd}(p+1,p^3+1)=p+1$, there exist integers $a,b$ such that
 \[
 a(p+1)+b(p^3+1)=p+1
 \]
 that is equivalent to $({\theta_1\over \theta_2})^{p+1}=({\theta_1\over \theta_2})^{a(p+1)+b(p^3+1)}=1$. It is easy to check that $p+1$ is a factor of $p^6-1$, so if there exist solutions, the number of solutions is $p+1$. It can be checked that there exist solutions in
$\mathbb{F}_{p}$ of equation system (\ref{ES01}). Finally, $T_3=p+1$ and $M_3=M_2+(p^m-1)T_3=(p+1)(q-1)+1$.
\end{IEEEproof}

 Using the following  notations, Lemma \ref{ESL001} can be restated in Lemma \ref{ES002} when $m$ is even.
Corresponding to Lemma \ref{FL002} and Corollary \ref{RQ02}, we introduce the following notations for convenience.
Let
\begin{equation}\label{0001}
N_{\varepsilon,j}=\left\{(\alpha,\beta,\gamma)\in \mathbb{F}_q^3\backslash\{(0,0,0)\}|S(\alpha,\beta,\gamma)=\varepsilon p^{{m+j}\over 2}\right\}
\end{equation}
where $\varepsilon=\pm 1$ and $j=0,2,4$. Also, denote $n_{\varepsilon,j}=|N_{\varepsilon,j}|$ for $j=0,2,4$.
And
\begin{equation}\label{0002}
N_{\varepsilon,j}=\left\{(\alpha,\beta,\gamma)\in \mathbb{F}_q^3\backslash\{(0,0,0)\}|S(\alpha,\beta,\gamma)=\varepsilon i p^{{m+j}\over 2}\right\}
\end{equation}
for $j=1,3$, where $i$ is the imaginary unit.
 By Lemma \ref{R001}, set
 \[
 n_{j}=n_{\varepsilon,j}=|N_{\varepsilon,j}|
 \]
  for $j=1,3$, since $m-j$ is odd.

 \begin{lemma}\label{ES002}
Let $p$ be an odd prime satisfying $p\equiv 3 \ \mbox{mod}\ 4$, and $q=p^m$ where $m$ is an even integer. Then the notations defined in (\ref{0001}) and (\ref{0002}) satisfy the following equations
\[
\begin{array}{lll}
 &&2(n_1+n_3)+n_{-1,0}+n_{1,0}+n_{-1,2}+n_{1,2}\\
 &&+n_{1,4}+n_{-1,4}\\
 & =&p^{3m}-1,\\
 && n_{1,0}-n_{-1,0}+p(n_{1,2}-n_{-1,2}) +p^2(n_{1,4}-n_{-1,4}) \\
   &=&p^{{m}\over 2}\left(p^{2m}-1\right), \\
 &&-2\left(pn_1+p^3n_3\right)
  +n_{1,0}+n_{-1,0}+p^2(n_{1,2}+n_{-1,2})\\
 &&+p^{4}(n_{1,4}+n_{-1,4}) \\
&=&p^{m} \left(p^m-1\right),\\
  && n_{1,0}-n_{-1,0}+p^3(n_{1,2}-n_{-1,2})+p^6(n_{1,4}-n_{-1,4}) \\
   &=& (p+1)p^{{3m}\over 2}\left(p^{m}-1\right).
\end{array}
\]
\end{lemma}

\begin{IEEEproof}
Substituting the symbols of (\ref{0001}) and (\ref{0002}) to Lemma \ref{ESL001}, we have the following four equations
 \[
\begin{array}{lll}
 &&2(n_1+n_3)+n_{-1,0}+n_{1,0}+n_{-1,2}+n_{1,2}\\
 &&+n_{1,4}+n_{-1,4}\\
 &=&p^{3m}-1,
\\
&&\sum\limits_{\alpha,\beta,\gamma\in \mathbb{F}_q} S(\alpha,\beta,\gamma)\\
&=&{p^{m\over 2}}(n_{1,0}-n_{-1,0})+ip^{{m+1}\over 2}(n_{1,1}-n_{-1,1})\\
&+&{p^{{m+2}\over 2}}(n_{1,2}-n_{-1,2})+ip^{{m+3}\over 2}(n_{1,3}-n_{-1,3})\\
&+&{p^{{m+4}\over 2}}(n_{1,4}-n_{-1,4})+p^m\\
&=&{p^{m\over 2}}(n_{1,0}-n_{-1,0})+{p^{{m+2}\over 2}}(n_{1,2}-n_{-1,2})\\
& +&{p^{{m+4}\over 2}}(n_{1,4}-n_{-1,4})+p^m\\
&=&p^{3m},
\end{array}
\]
 \[
\begin{array}{lll}
&&\sum\limits_{\alpha,\beta,\gamma\in \mathbb{F}_q} S(\alpha,\beta,\gamma)^2\\
&=&{p^{m}}(n_{1,0}+n_{-1,0})-p^{{m+1} }(n_{1,1}+n_{-1,1})\\
&&{p^{{m+2} }}(n_{1,2}+n_{-1,2})-p^{{m+3} }(n_{1,3}+n_{-1,3})\\
&+&{p^{{m+4}}}(n_{1,4}+n_{-1,4})+p^{2m}\\
&=&p^{3m}, \\
 &&\sum\limits_{\alpha,\beta,\gamma\in \mathbb{F}_q} S(\alpha,\beta,\gamma)^3\\
&=&{p^{{3m}\over 2}}(n_{1,0}-n_{-1,0})-ip^{{3(m+1)}\over 2}(n_{1,1}-n_{-1,1})\\
& +&{p^{{3(m+2)}\over 2}}(n_{1,2}-n_{-1,2})-ip^{{3(m+3)}\over 2}(n_{1,3}-n_{-1,3})\\
&+&{p^{{3(m+4)}\over 2}}(n_{1,4}-n_{-1,4})+p^{3m}\\
&=& {p^{{3m}\over 2}}(n_{1,0}-n_{-1,0})+{p^{{3(m+2)}\over 2}}(n_{1,2}-n_{-1,2})\\
&+&{p^{{3(m+4)}\over 2}}(n_{1,4}-n_{-1,4})+p^{3m}\\
&=&\left((p+1)\left(p^m-1\right)+1\right)p^{3m}
\end{array}
\]
 where the first one comes from the fact that there are $p^{3m}-1$ elements in the set $\mathbb{F}_q^3\backslash\{(0,0,0)\}$. Also, note that $S(\alpha,\beta,\gamma)=p^m$ when $\alpha=\beta=\gamma=0$.

Using $n_{j}=n_{\varepsilon,j}=|N_{\varepsilon,j}|$ for $j=1,3$, the result is obtained by simplification.
\end{IEEEproof}

\subsection{The Fourth Moment of $S(\alpha,\beta,\gamma)$}\label{Sec3.2}


For the fourth moment of $S(\alpha,\beta,\gamma)$ in the particular case of $p=3$, there is the
following result about the number of solutions of the equation system
\begin{equation}\label{E02}
\left\{
\begin{array}{ll}
x^2+y^2+z^2+1 &=0\\
x^{p+1}+y^{p+1}+z^{p+1}+1 &=0\\
x^{p^2+1}+y^{p^2+1}+z^{p^2+1}+1 &=0
\end{array}
\right.
\end{equation}
in Lemma \ref{CM001}, which is denoted by $T_4$.

\begin{lemma}(\cite{LL002})\label{CM001}
Let $p=3$ and $q=p^m$, then
\[
T_4=4\left(2p^m-3\right).
\]
\end{lemma}

Using Lemma \ref{CM001} and $M_3$ in Lemma \ref{ESL001}, $M_4$ is calculated in Lemma \ref{CM01} where $M_4=M_3+(q-1)T_4$.
\begin{lemma}\label{CM01}
Let $p=3$ and $q=p^m$. The number of solutions of the following equation system
\begin{equation}\label{E01}
\left\{
\begin{array}{ll}
x^2+y^2+z^2+w^2 &=0\\
x^{p+1}+y^{p+1}+z^{p+1}+w^{p+1} &=0\\
x^{p^2+1}+y^{p^2+1}+z^{p^2+1}+w^{p^2+1} &=0\\
\end{array}
\right.
\end{equation}
is $M_4=8\left(p^m-1\right)^2+1$.
\end{lemma}

Corresponding to Lemma \ref{ESL001}, the result of the fourth moment is provided in Lemma \ref{S04} by applying Lemma \ref{CM01}. \begin{lemma}\label{S04}
Let $p=3$ and $q=p^m$. Then
\[
\sum\limits_{\alpha,\beta,\gamma\in \mathbb{F}_q} S(\alpha,\beta,\gamma)^4=M_4\cdot p^{3m}=\left(8(p^m-1)^2+1\right)p^{3m}.
\]
\end{lemma}

Corresponding to Lemma \ref{ES002}, Lemma \ref{S04} can be rewritten as the following corollary using the symbols of (\ref{0001}) and (\ref{0002}).
\begin{corollary}\label{C02}
Let $p=3$ and $q=p^m$ where $m$ is an even integer. Then
\[
\begin{array}{lll}
&&n_{1,0}+n_{-1,0}+2n_1p^2+p^4(n_{1,2}+n_{-1,2})+2n_3p^6\\
  &+&p^8(n_{1,4}+n_{-1,4})\\
&=&\left(8(p^m-1)^2-p^m+1\right)p^m.
\end{array}
\]
\end{corollary}

\subsection{The Fifth Moment of $S(\alpha,\beta,\gamma)$}\label{Sec3.3}

  For the fifth moment of $S(\alpha,\beta,\gamma)$, we need Magma \cite{BC001} to find the number of solutions of the following equation system
\begin{equation}\label{E03}
\left\{
\begin{array}{ll}
x^2+y^2+z^2+w^2+u^2 &=0\\
x^{p+1}+y^{p+1}+z^{p+1}+w^{p+1}+u^{p+1} &=0\\
x^{p^2+1}+y^{p^2+1}+z^{p^2+1}+w^{p^2+1}+u^{p^2+1} &=0\\
\end{array}
\right.
\end{equation}
which is denoted by $M_5$.

 As in \cite{BM001}, the irreducible components corresponding to the projective variety defined by (\ref{E03}) are listed in Table \ref{TAB02} using Magma.
It is easy to be verified that every block of Table \ref{TAB02} contains a system of three equations (note that `$=0$' is omitted), the solutions of which satisfy (\ref{E03}). Furthermore, the union of all the solutions in each block presents the solutions of (\ref{E03}) exactly. Those equation systems are circulant symmetric about the variables. In general, few works were provided to deal with the moments using five variables. But in this paper, Magma helps us on the reduction of such systems in Lemma \ref{CM02}, Lemma \ref{S004} and Corollary \ref{C03}. For relevant knowledge of algebraic geometry, the reader is referred to \cite{S001}.

\begin{table*}[htbp]
 \renewcommand\thetable{\Roman{table}}
 \caption{ }\label{TAB02}
 \hspace{2cm}
 \begin{tabular}{|l|l|l|l|l|l|l|l|}
\hline
$x^2$, & $x^2$, & $x^2$, & $x^2$,  & $x^2$, & $x^2$, & $x^2$, & $x^2$, \\
    $y -w -u$, &  $y -w -u$, &  $y -w + u$, &  $y -w + u$, & $y + w -u$, & $y + w -u$, & $y + w + u$,& $y + w + u$, \\
    $z -w + u$, &  $z + w -u$, & $z -w -u$, &  $z + w + u$, & $z -w -u$, & $z +w + u$, &  $z -w + u$, & $z + w -u$,\\
\hline
$y^2$, & $y^2$, & $y^2$, & $y^2$,  & $y^2$, & $y^2$, & $y^2$, & $y^2$, \\
    $x-w -u$, &  $x -w -u$, &  $x -w + u$, &  $x -w + u$, & $x + w -u$, & $x + w -u$, & $x + w + u$,& $x + w + u$, \\
    $z -w + u$, & $z + w -u$, & $z -w -u$, & $z + w + u$, & $z -w -u$, & $z +w + u$, & $z -w + u$, & $z + w -u$,\\
\hline
$z^2$, & $z^2$, & $z^2$, & $z^2$,  & $z^2$, & $z^2$, & $z^2$, & $z^2$, \\
   $x-w -u$, &   $x -w -u$, &   $x -w + u$, &  $x -w + u$, & $x + w -u$, & $x + w -u$, & $x + w + u$,& $x + w + u$, \\
    $y -w + u$, & $y + w -u$, & $y -w -u$, &  $y + w + u$, & $y -w -u$, & $y +w + u$, &  $y -w + u$, & $y + w -u$,\\
\hline
$w^2$, & $w^2$, & $w^2$, & $w^2$,  & $w^2$, & $w^2$, & $w^2$, & $w^2$, \\
   $x-z -u$, &   $x -z -u$, &   $x -z + u$, &  $x -z + u$, & $x + z -u$, & $x + z -u$, & $x + z+ u$,& $x + z + u$, \\
    $y -z + u$, &  $y + z -u$, & $y -z -u$, &  $y + z + u$, & $y -z -u$, & $y +z + u$, &  $y -z + u$, & $y + z -u$,\\
\hline
$u^2$, & $u^2$, & $u^2$, & $u^2$,  & $u^2$, & $u^2$, & $u^2$, & $u^2$, \\
   $x-z -w$, &   $x -z -w$, &   $x -z + w$, &  $x -z + w$, & $x + z -w$, & $x + z -w$, & $x + z+ w$,& $x + z + w$, \\
    $y -z + w$, &  $y + z -w$, & $y -z -w$, &  $y + z + w$, & $y -z -w$, & $y +z + w$, &  $y -z + w$, & $y + z -w$,\\
\hline

\end{tabular}
\end{table*}

\begin{multicols}{3}{
Put the text here. Maths, tabulars, pictures etc are all ok (but not
figures and tables). Remember to load in the \texttt{multicol}
package at the top of your document.
}
\end{multicols}

\begin{lemma}\label{CM02}
Let $p=3$ and $q=p^m$. Then
  \[
  M_5=5\left(p^m-1\right)\left(8p^m-2p-10\right)+1.
\]
\end{lemma}

\begin{IEEEproof}
According to the proof of Lemma \ref{ESL001}, the number of nonzero solutions of (\ref{ES201}) is $M_2'=M_2-1=0$, and the number of nonzero elements in (\ref{CM0001}) is $M_3'=(p+1)(q-1)$. By Lemma \ref{CM01}, the number of nonzero solutions of (\ref{E01}) is $M_4'=M_4-4M_3'-1= \left(p^m-1\right)\left(8p^m-4p-12\right)$.

For the solutions of equation system (\ref{E03}), by Table \ref{TAB02} we find that at least one of the elements $x,y,z,w,u$ is zero, and there are two cases to be considered.
\begin{itemize}
\item If only one of the five variables is zero, the number of such solutions is
\[
5M_4'=5\left(p^m-1\right)\left(8p^m-4p-12\right).
\]
\item If two variables are zero, the number of such solutions is
\[
\left(
\begin{array}{c}
5\\
2
\end{array}
\right)M_3'=10(p+1)(q-1).
\]
\end{itemize}
Altogether, the number of solutions of equation system (\ref{E03}) is
\[
5M_4'+10M_3'+1=5\left(p^m-1\right)\left(8p^m-2p-10\right)+1.
\]
\end{IEEEproof}

Applying Lemma \ref{CM02}, the result about the fifth moment of exponential sum $S(\alpha,\beta,\gamma)$ is obtained.
\begin{lemma}\label{S004}
Let $p=3$ and $q=p^m$. Then
\[
\begin{array}{lll}
&&\sum\limits_{\alpha,\beta,\gamma\in \mathbb{F}_q} S(\alpha,\beta,\gamma)^5\\
&=&M_5\cdot p^{3m}\\
&=&\left(5(p^m-1)(8p^m-2p-10)+1\right)p^{3m}.
\end{array}
\]
\end{lemma}

Using the symbols of (\ref{0001}) and (\ref{0002}), Lemma \ref{S004} can be rewritten as Corollary \ref{C03}.
\begin{corollary}\label{C03}
Let $p=3$ and $q=p^m$ where $m$ is an even integer. Then
\[
\begin{array}{lll}
&&n_{1,0}-n_{-1,0}+p^5(n_{1,2}-n_{-1,2})+ p^{10}(n_{1,4}-n_{-1,4})\\
&=&\left(5(p^m-1)(8p^m-2p-10)-p^{2m}+1\right)p^{m\over 2}.
\end{array}
\]
\end{corollary}

\subsection{MacWilliams' Identities}  \label{Sec3.4}

 MacWilliams' theorem is for Hamming weight enumerators of linear codes over finite field $\mathbb{F}_p$ \cite {MS001}. Using this theorem, Lemma \ref{SM02} is provided for the weight distribution using dual code's weight distribution of Lemma \ref{W001}. The two identities in Lemma \ref{SM02}
 will combine with previous identities in final result.

Let $A_i$ be the number of codewords of weight $i$ in a code $C$ with length $l$ and dimension $k$ where $0\leq i\leq l$. Let $A_i'$ be the corresponding number in the dual code $C^{\perp}$. Then
 \begin{equation}\label{MS01}
W_C(x,y)={1\over |C^{\perp}|}W_{C^{\perp}}(x+(p-1)y,x-y)
\end{equation}
where $W_C(x,y) = \sum\limits_{i=0}^{l}{A_ix^{l-i}y^i}$. Setting $x=1$, equation (\ref{MS01}) changes to
\begin{equation}\label{MS002}
\sum\limits_{i=0}^{l}A_iy^i = {1\over {p^{l-k}}}\sum\limits_{i=0}^{l}A_i'(1+(p-1)y)^{l-i}(1-y)^i.
\end{equation}
After differentiating (\ref{MS002}) with respect to $y$, we have
\[
\begin{array}{lll}
&&\sum\limits_{i=1}^{l}iA_iy^{i-1}\\
&=&{1\over {p^{l-k}}}\sum\limits_{i=0}^{l}A_i'\{(l-i)(p-1)(1+(p-1)y)^{l-i-1}(1-y)^i\\
&&+(1+(p-1)y)^{l-i}i(-1)(1-y)^{i-1}\}.
\end{array}
\]
 Setting $y=1$, the first MacWilliams' moment identity is obtained for $l\geq 2$
\[
\sum\limits_{i=1}^{l}{{iA_i}\over{p^k}}={1\over p}((p-1)l-A_1')={1\over p}(p-1)l \ \ \ \mbox{if} \ \ A_1'=0.
\]
Differentiating again,
\[
\begin{array}{lll}
&&\sum\limits_{i=1}^{l}i(i-1)A_iy^{i-2} \\
&=&{1\over {p^{l-k}}}\sum\limits_{i=0}^{l}A_i'\{(l-i)(l-i-1)(p-1)^2(1+(p-1)y)^{l-i-2}\\
&&(1-y)^i+2(l-i)(p-1)(1+(p-1)y)^{l-i-1}i(-1)\\
&&(1-y)^{i-1}+(1+(p-1)y)^{l-i}i(i-1)(1-y)^{i-2}  \}.
\end{array}
\]
Substituting $y=1$, the second MacWilliams' moment identity is obtained
\begin{equation}\label{SM001}
\sum\limits_{i=1}^{l}i(i-1)A_i={1\over {p^{l-k}}}\left(l(l-1)(p-1)^2p^{l-2}+2A_2'p^{l-2}\right)\ \ \ \mbox{if} \ \ A_1'=0.
\end{equation}
 Differentiating for the third and fourth time,
\[
 \begin{array}{lll}
 &&\sum\limits_{i=1}^{l}i(i-1)(i-2)A_iy^{i-3} \\
 &=&{1\over {p^{l-k}}}\sum\limits_{i=0}^{l}A_i'\{(l-i)(1+(p-1)y)^{l-i-3}(1-y)^i(p-1)^3\\
 &&(l-i-1)(l-i-2)+3(l-i)(l-i-1)(p-1)^2\\
 &&(1+(p-1)y)^{l-i-2}(1-y)^{i-1}i(-1)\\
  &+&3(l-i)(p-1)(1+(p-1)y)^{l-i-1}(1-y)^{i-2}i(i-1)  \\
  &+&(1+(p-1)y)^{l-i}i(i-1)(i-2)(-1)(1-y)^{i-3}\}
 \end{array}
\]
and
\[
\begin{array}{lll}
&&\sum\limits_{i=1}^{l}i(i-1)(i-2)(i-3)A_iy^{i-4} \\
&=&{1\over {p^{l-k}}}\sum\limits_{i=0}^{l}A_i'\{(l-i)(1+(p-1)y)^{l-i-4}(1-y)^i(p-1)^4\\
&&(l-i-1)(l-i-2)(l-i-3)+4(l-i)\\
&&(1+(p-1)y)^{l-i-3}i(-1)(1-y)^{i-1}(p-1)^3(l-i-1)\\
&&(l-i-2)+6(l-i)(l-i-1)(p-1)^2\\
&&(1+(p-1)y)^{l-i-2}(1-y)^{i-2}i(i-1)\\
&+&4(l-i)(p-1)(1+(p-1)y)^{l-i-1}\\
&&(1-y)^{i-3}i(i-1)(i-2)(-1)+(1+(p-1)y)^{l-i}\\
&&i(i-1)(i-2)(i-3)(1-y)^{i-4}\}.
\end{array}
\]
Substituting $y=1$, if $A_1'=A_3'=0$, the fourth MacWilliams' moment identity is obtained
\begin{equation}\label{SM003}
\begin{array}{lll}
&&\sum\limits_{i=1}^{l}i(i-1)(i-2)(i-3)A_i  \\
&=& {1\over {p^{l-k}}}\{l(l-1)(l-2)(l-3)p^{l-4}(p-1)^4\\
&+&12A_2'(l-2)(l-3)(p-1)^2p^{l-4}+24A_4'p^{l-4}\}.
 \end{array}
\end{equation}

\begin{lemma}\label{W001}
Let $p=3, q=p^m$. Let $\mathcal{C}$ denote the cyclic code with nonzeros $\pi^{-2}, \pi^{-(p+1)}$ and $\pi^{-(p^2+1)}$, the weights of the dual code $\mathcal{C}^{\perp}$ satisfy the following
\[
A_0'=1, \ A_1'=0, \ A_2'=p^m-1,\ A_3'=0,
\]
\[
 A_4'={{\left(p^m-1\right)(2p^m-p-3)}\over 3}+{{(p^{m}-1)(p^{m}-3)}\over 2}.
\]
\end{lemma}

\begin{IEEEproof}
Below, codewords are considered in the dual code.
Easy to see that $A_0'=1$ and $A_1'=0$. For the codewords with weight two, if the components at the two positions have the same value, by equation (\ref{ES201}) we find that $M_2'=0$. Let's consider the following equation system about the positions
\[
\left\{
 \begin{array}{ll}
 x^2-y^2 &=0\\
 x^{p+1}-y^{p+1} &=0\\
 x^{p^2+1}-y^{p^2+1} &=0,
 \end{array}
\right.
\]
which should be satisfied by the coordinates of the codewords.
It can be checked that for any $y\in \mathbb{F}_q^*$, $x=-y$ is the other corresponding coordinate. That is $A_2'=p^m-1$.

As to weight-three codewords, there are two cases to be considered.
 \begin{enumerate}
 \renewcommand{\labelenumi}{$($\mbox{\roman{enumi}}$)$}
 \item  If all the values corresponding to the three coordinates of the codeword are the same, it is necessary to study the solutions of the following equation system
\begin{equation}\label{ES0001}
\left\{
\begin{array}{ll}
x^2+y^2+1 &=0 \\
x^{p+1}+y^{p+1}+1 &=0 \\
x^{p^2+1}+y^{p^2+1}+1 &=0,
\end{array}
\right.
\end{equation}
which should be satisfied by the coordinates of the codewords.
From the first two equations of (\ref{ES0001}), we find that $x^2=y^2=1$ contradicting the fact that $x,y,1$ should be different.
\item If one value is different from the other two values at the three coordinates, consider
\begin{equation}\label{ES001}
\left\{
\begin{array}{ll}
x^2-y^2+1 &=0 \\
x^{p+1}-y^{p+1}+1 &=0 \\
x^{p^2+1}-y^{p^2+1}+1 &=0.
\end{array}
\right.
\end{equation}
Solving the above system, we have $x=0$ contradicting the fact that the coordinates should be different from $0$.
\end{enumerate}
Combing the above two cases, $A_3'=0$.

Now, let's consider the number of codewords with weight four in three cases.
\begin{enumerate}
 \renewcommand{\labelenumi}{$($\mbox{\roman{enumi}}$)$}
 \item Case I: at the four positions, the components have the same value. According to the proof of Lemma \ref{CM02}, we know that the number of nonzero solutions of equation system (\ref{E01}) is $M_4'=\left(p^m-1\right)\left(8p^m-4p-12\right)$. For a solution $(x,y,z,w)$ of (\ref{E01}), if two of them are equal, e.g., $z=w=v$, then (\ref{E01}) becomes
\begin{equation}\label{ES003}
\left\{
\begin{array}{ll}
x^2+y^2-v^2 &=0 \\
x^{p+1}+y^{p+1}-v^{p+1} &=0 \\
x^{p^2+1}+y^{p^2+1}-v^{p^2+1} &=0.
\end{array}
\right.
\end{equation}
Solving the above system, it can be found that $x$ or $y$ is zero, so the number of nonzero solutions of (\ref{ES003}) is $0$. Then all those $M_4'$ nonzero solutions of (\ref{E01}) correspond to the codewords where $24=4!$ solutions correspond to a four-tuple and each tuple corresponds to two codewords.
 Therefore, there are $2\cdot {{M_4'}\over {24}}={{M_4'}\over {12}}$ codewords in this case.
\item Case II: one value at the four nonzero positions are different from the other three values. Then it is necesssary to consider the solutions of the following system
    \begin{equation}\label{E002}
\left\{
\begin{array}{ll}
x^2+y^2+z^2-w^2 &=0\\
x^{p+1}+y^{p+1}+z^{p+1}-w^{p+1} &=0\\
x^{p^2+1}+y^{p^2+1}+z^{p^2+1}-w^{p^2+1} &=0.
\end{array}
\right.
\end{equation}
 Using Mamga \cite{BC001}, the irreducible components of the projective variety corresponding to (\ref{E002}) are provided by the polynomials listed in Table \ref{TAB001}. Easy to see that at least one of $x,y,z,w$ is zero, so the solutions can not correspond to codewords.
\begin{table*}[htbp]
 \renewcommand\thetable{\Roman{table}}
 \caption{ }\label{TAB001}
 \hspace{3cm}
 \begin{tabular}{|l|l|l|l|l|l|}
\hline
$y^4$, &   $y^4$,     & $z^4$,       & $z^4$,      & $z^4$,  \\
$x^2+y^2$, &$x^2+y^2$, &  $x^2+z^2$, &  $x^2+z^2$, & $y^2+z^2$,  \\
$z -w $,   & $z + w $, & $y -w  $,    &  $y + w $,    & $x-w$, \\
\hline
$z^4$,          & $w^4$,            & $w^4$,             & $w^4$,              & $w^4$,  \\
$y^2+z^2$,      &  $y^2-yz+z^2+w^2$,  & $y^2-yz+z^2+w^2$,  &  $y^2+yz+z^2+w^2$, & $y^2+yz+z^2+w^2$, \\
 $x+w$,           & $x-y+z$,          & $x+y-z$,           & $x-y-z$,      & $x+y+z$, \\
\hline

\end{tabular}
\end{table*}
\item Case III, two values at the coordinates are the same. Let's consider the number of solutions of the following equation system
   \begin{equation}\label{E003}
\left\{
\begin{array}{ll}
x^2+y^2-z^2-w^2 &=0\\
x^{p+1}+y^{p+1}-z^{p+1}-w^{p+1} &=0\\
x^{p^2+1}+y^{p^2+1}-z^{p^2+1}-w^{p^2+1} &=0.
\end{array}
\right.
\end{equation}
Again the irreducible components are presented by the polynomials listed in Table \ref{TAB003},
\begin{table*}[htbp]
 \renewcommand\thetable{\Roman{table}}
 \caption{ }\label{TAB003}
 \hspace{4.1cm}
 \begin{tabular}{|l|l|l|l|l|l|l|l|}
\hline
    $x-z$, &  $x-z$, &  $x+z$, &  $x+z$, & $x-w$, & $x-w$, & $x+w$,& $x+w$, \\
    $y-w$, &  $y+w$, & $y-w$, &  $y+w$, & $y-z$, & $y+z$, &  $y-z$, & $y+z$,\\
\hline

\end{tabular}
\end{table*}
by which only the cases $x=-z,y=-w$ and $x=-w,y=-z$ are the possible solutions which can correspond to codewords since coordinates should be different. And the number of such solutions is $2(p^{m}-1)(p^{m}-3)$ which corresponds to ${4\cdot {{2(p^{m}-1)(p^{m}-3)}\over 16}}={{(p^{m}-1)(p^{m}-3)}\over 2}$ codewords, since every four-tuple $(x_0,y_0,z_0,w_0)$ corresponds to $4\cdot 2 \cdot 2$ solutions of (\ref{E003}).
 In fact, if $c$ is a weight-four codeword with nonzero positions and values $(x_0,y_0,-x_0,-y_0)\longrightarrow(1,1,-1,-1)$, then $(x_0,y_0,-x_0,-y_0)\longrightarrow(-1,-1,1,1)$, $(x_0,y_0,-x_0,-y_0)\longrightarrow(-1,1,1,-1)$ and $(x_0,y_0,-x_0,-y_0)\longrightarrow(1,-1,-1,1)$ can all represent weight-four codewords.
 \end{enumerate}
Combing the above three cases,
\[
\begin{array}{ll}
A_4'&={{M_4'}\over {12}}+{{(p^{m}-1)(p^{m}-3)}\over 2} \\
    &={{\left(p^m-1\right)(2p^m-p-3)}\over 3}+{{(p^{m}-1)(p^{m}-3)}\over 2}.
    \end{array}
\]
The result of the lemma is obtained.
\end{IEEEproof}

\begin{lemma} \label{SM02}
Let $p=3, q=p^m$ where $m\geq 6$ is an even integer. The notations defined in equations (\ref{0001}) and (\ref{0002}) satisfy the following equations
\begin{equation}\label{SM004}
\begin{array}{lll}
 n_{1,0}+n_{-1,0}+p^2(n_{1,2}+n_{-1,2})+p^4(n_{1,4}+n_{-1,4})&=a\\
 n_{1,0}+n_{-1,0}+p^4(n_{1,2}+n_{-1,2})+p^8(n_{1,4}+n_{-1,4})&=b
\end{array}
\end{equation}
where
\[
\begin{array}{lll}
a&=&\{p^{3m-2}\left((p^m-1)(p^m-2)(p-1)^2+2A_2'\right)\\
&-&p^{2(m-1)}(p-1)^2(p^{3m}-2p^{2m}+1)\\
   &+&(p-1)(p^m-1)p^{3m-1}\}/\left((p-1)^2p^{m-2}\right),
\end{array}
\]
\[
\begin{array}{lll}
 b&=&\{p^{3m-4}\{(p^m-1)(p^m-2)(p^m-3)(p^m-4)(p-1)^4\\
                  &+&12A_2'(p^m-3)(p^m-4)(p-1)^2+24A_4'\}\\
                  &-&(p-1)^5p^{3(m-1)}a\\
                  &-&p^{4(m-1)}(p-1)^4(p^{3m}-4p^{2m}-16p^m+19)\\
                  &+&6\{(p-1)^3p^{2(m-1)}a\\
                  &+&p^{3(m-1)}(p-1)^3(p^{3m}-p^{2m+1}-4p^m+6)\}\\
                  &-&11\{(p-1)^2p^{m-2}a\\
                  &+&p^{2(m-1)}(p-1)^2(p^{3m}-2p^{2m}+1)\}\\
                  &+&6\left((p-1)(p^m-1)p^{3m-1}\right)\}/\left((p-1)^4p^{2m-4}\right),
 \end{array}
\]
 and $A_2', A_4'$ are defined in Lemma \ref{W001}.
\end{lemma}

\begin{IEEEproof}
Define the following notations for simplification
\begin{equation}\label{P001}
\begin{array}{lll}
&R_{00}=p^{m-1}(p-1), R_0={{p-1}\over p}{p^{m\over 2}}, R_2 ={{p-1}\over p}{p^{{m+2}\over 2}}, \\
&R_4 ={{p-1}\over p}{p^{{m+4}\over 2}}.
\end{array}
\end{equation}
In addition, the usage of MacWilliams identities in the following paragraph, implies the condition $m\geq 6$, refer to Lemma \ref{GCD002}. By equation (\ref{C01}) and Lemma \ref{ES02}, $\mathcal{C}$ has seven possible nonzero weights
\[
\begin{array}{lll}
&A_{R_{00}}=2(n_1+n_3), A_{R_{00}-R_0}=n_{1,0}, A_{R_{00}+R_0}=n_{-1,0},\\
 &A_{R_{00}-R_2}=n_{1,2}, A_{R_{00}+R_2}=n_{-1,2}, A_{R_{00}-R_4}=n_{1,4}, \\
 &A_{R_{00}+R_4}=n_{-1,4}.
\end{array}
\]
With the above notations, the first four moments of codeword weight can be computed
\begin{equation}\label{FMI001}
\begin{array}{lll}
\sum\limits_{i=0}^{l}iA_i&=&R_{00}2(n_1+n_3)+ (R_{00}-R_0)n_{1,0}\\
                          &+& (R_{00}+R_0)n_{-1,0}+(R_{00}-R_2)n_{1,2}\\
                          &+&(R_{00}+R_2)n_{-1,2} +(R_{00}-R_4)n_{1,4}\\
                        &+&(R_{00}+R_4)n_{-1,4}\\
                        &=&R_{00}\{2(n_1+n_3)+n_{1,0}+n_{-1,0}+n_{1,2}+n_{-1,2}\\
                        &+&n_{1,4}+n_{-1,4}\}-(R_0(n_{1,0}-n_{-1,0})\\
                        &+&R_2(n_{1,2}-n_{-1,2})+R_4(n_{1,4}-n_{-1,4}))\\
                        &=&(p^{3m}-1)p^{m-1}(p-1)-{{p-1}\over p}p^{m\over 2}(p^{m\over 2}(p^{2m}-1))\\
                        &=&(p-1)(p^m-1)p^{3m-1},
\end{array}
\end{equation}
\begin{equation}\label{SMI001}
\begin{array}{lll}
\sum\limits_{i=0}^{l}i^2A_i&=&R_{00}^22(n_1+n_3)+ (R_{00}-R_0)^2n_{1,0}\\
                           &+& (R_{00}+R_0)^2n_{-1,0}+(R_{00}-R_2)^2n_{1,2}\\
                           &+&(R_{00}+R_2)^2n_{-1,2} +(R_{00}-R_4)^2n_{1,4}\\
                           &+&(R_{00}+R_4)^2n_{-1,4}\\
                        &=&R_{00}^2\{2(n_1+n_3)+n_{1,0}+n_{-1,0}+n_{1,2}+n_{-1,2}\\
                        &+&n_{1,4}+n_{-1,4}\}-2R_{00}\{R_0(n_{1,0}-n_{-1,0})\\
                        &+&R_2(n_{1,2}-n_{-1,2})+R_4(n_{1,4}-n_{-1,4})\}\\
                        &+&R_{0}^2(n_{1,0}+n_{-1,0})+R_{2}^2(n_{1,2}+n_{-1,2})\\
                        &+&R_{4}^2(n_{1,4}+n_{-1,4})\\
                        &=&(p-1)^2p^{m-2}(n_{1,0}+n_{-1,0}+p^2(n_{1,2}+n_{-1,2})\\
                        &+&p^4(n_{1,4}+n_{-1,4})) \\
                        &+&p^{2(m-1)}(p-1)^2(p^{3m}-2p^{2m}+1),
\end{array}
\end{equation}
\begin{equation}\label{TMI001}
\begin{array}{lll}
\sum\limits_{i=0}^{l}i^3A_i&=&R_{00}^32(n_1+n_3)+ (R_{00}-R_0)^3n_{1,0}\\
                            &+& (R_{00}+R_0)^3n_{-1,0}+(R_{00}-R_2)^3n_{1,2}\\
                            &+&(R_{00}+R_2)^3n_{-1,2} +(R_{00}-R_4)^3n_{1,4}\\
                            &+&(R_{00}+R_4)^3n_{-1,4}\\
                        &=&R_{00}^3\{2(n_1+n_3)+n_{1,0}+n_{-1,0}+n_{1,2}+n_{-1,2}\\
                        &+&n_{1,4}+n_{-1,4}\}-3R_{00}^2\{R_0(n_{1,0}-n_{-1,0})\\
                        &+&R_2(n_{1,2}-n_{-1,2})+R_4(n_{1,4}-n_{-1,4})\}\\
                        &+&3R_{00}\{R_{0}^2(n_{1,0}+n_{-1,0})+R_{2}^2(n_{1,2}+n_{-1,2})\\
                        &+&R_{4}^2(n_{1,4}+n_{-1,4})\}-\{R_0^3(n_{1,0}-n_{-1,0})\\
                        &+&R_2^3(n_{1,2}-n_{-1,2})+R_4^3(n_{1,4}-n_{-1,4})\}\\
                        &=&(p-1)^3p^{2(m-1)}\{n_{1,0}+n_{-1,0}\\
                        &+&p^2(n_{1,2}+n_{-1,2})+p^4(n_{1,4}+n_{-1,4})\} \\
                        &+&p^{3(m-1)}(p-1)^3(p^{3m}-p^{2m+1}-4p^m+6),
\end{array}
\end{equation}
\begin{equation}\label{FMI002}
\begin{array}{lll}
\sum\limits_{i=0}^{l}i^4A_i&=&R_{00}^42(n_1+n_3)+ (R_{00}-R_0)^4n_{1,0}\\
                           &+& (R_{00}+R_0)^4n_{-1,0}+(R_{00}-R_2)^4n_{1,2}\\
                           &+&(R_{00}+R_2)^4n_{-1,2} +(R_{00}-R_4)^4n_{1,4}\\
                           &+&(R_{00}+R_4)^4n_{-1,4}\\
                        &=&R_{00}^4\{2(n_1+n_3)+n_{1,0}+n_{-1,0}+n_{1,2}+n_{-1,2}\\
                        &+&n_{1,4}+n_{-1,4}\}-4R_{00}^3\{R_0(n_{1,0}-n_{-1,0})\\
                        &+&R_2(n_{1,2}-n_{-1,2})+R_4(n_{1,4}-n_{-1,4})\}\\
                        &+&6R_{00}^2\{R_{0}^2(n_{1,0}+n_{-1,0})+R_{2}^2(n_{1,2}+n_{-1,2})\\
                        &+&R_{4}^2(n_{1,4}+n_{-1,4})\}-4R_{00}\{R_0^3(n_{1,0}-n_{-1,0})\\
                        &+&R_2^3(n_{1,2}-n_{-1,2})+R_4^3(n_{1,4}-n_{-1,4})\}\\
                        &+&(R_0^4(n_{1,0}+n_{-1,0})+R_2^4(n_{1,2}+n_{-1,2})\\
                        &+&R_4^4(n_{1,4}+n_{-1,4}))\\
                        &=&(p-1)^5p^{3(m-1)}\{n_{1,0}+n_{-1,0}\\
                        &+&p^2(n_{1,2}+n_{-1,2})+p^4(n_{1,4}+n_{-1,4})\} \\
                        &+&(p-1)^4p^{2m-4}\{n_{1,0}+n_{-1,0}\\
                        &+&p^4(n_{1,2}+n_{-1,2})+p^8(n_{1,4}+n_{-1,4})\}\\
                        &+&p^{4(m-1)}(p-1)^4(p^{3m}-4p^{2m}-16p^m+19).
\end{array}
\end{equation}
From equations (\ref{SM001}), (\ref{FMI001}) and (\ref{SMI001}) we have
\[
\begin{array}{lll}
\sum\limits_{i=1}^{l}i(i-1)A_i&=&{1\over {p^{l-k}}}\left(l(l-1)(p-1)^2p^{l-2}+2A_2'p^{l-2}\right)\\
                              &=&p^{3m-2}\left(l(l-1)(p-1)^2+2A_2'\right)\\
                              &=&\sum\limits_{i=1}^{l}i^2A_i-\sum\limits_{i=1}^{l}iA_i\\
                               &=&(p-1)^2p^{m-2}\{n_{1,0}+n_{-1,0}\\
                               &+&p^2(n_{1,2}+n_{-1,2})+p^4(n_{1,4}+n_{-1,4})\} \\
                             &+&p^{2(m-1)}(p-1)^2(p^{3m}-2p^{2m}+1)\\
                             &-&(p-1)(p^m-1)p^{3m-1},
\end{array}
\]
and the first one of equation (\ref{SM004}) is obtained after simplification.

According to the fourth moment of MacWilliams' moment identity (\ref{SM003}) and equations (\ref{FMI001}), (\ref{SMI001}), (\ref{TMI001}) and (\ref{FMI002}),
\[
\begin{array}{lll}
&& \sum\limits_{i=1}^{l}i(i-1)(i-2)(i-3)A_i \\
 &=& {1\over {p^{l-k}}}\{l(l-1)(l-2)(l-3)p^{l-4}(p-1)^4\\
 &+&12A_2'(l-2)(l-3)(p-1)^2p^{l-4}+24A_4'p^{l-4}\}\\
                                          &=&p^{3m-4}\{(p^m-1)(p^m-2)(p^m-3)(p^m-4)(p-1)^4\\
                                          &+&12A_2'(p^m-3)(p^m-4)(p-1)^2+24A_4'\}\\
                                          &=&\sum\limits_{i=1}^{l}i^4A_i-6\sum\limits_{i=1}^{l}i^3A_i+11\sum\limits_{i=1}^{l}i^2A_i-6\sum\limits_{i=1}^{l}iA_i\\
                                          &=& (p-1)^4p^{2m-4}\{n_{1,0}+n_{-1,0}+p^4(n_{1,2}+n_{-1,2})\\
                                          &+&p^8(n_{1,4}+n_{-1,4})\}+ (p-1)^5p^{3(m-1)}a\\
                                          &+&p^{4(m-1)}(p-1)^4(p^{3m}-4p^{2m}-16p^m+19)\\
                                          &-&6\{(p-1)^3p^{2(m-1)}a\\
                                          &+&p^{3(m-1)}(p-1)^3(p^{3m}-p^{2m+1}-4p^m+6)\}\\
                                          &+&11\left((p-1)^2p^{m-2}a+p^{2(m-1)}(p-1)^2(p^{3m}-2p^{2m}+1)\right)\\
                                          &-&6\left((p-1)(p^m-1)p^{3m-1}\right),
 \end{array}
\]
and the second one of equation (\ref{SM004}) is obtained after simplification.
\end{IEEEproof}

\subsection{Weight Distribution of $\mathcal{C}$}\label{Sec3.5}

In this subsection, the parameters defined in equations (\ref{0001}) and (\ref{0002}) are calculated in Lemma \ref{SM03}, and the weight distribution of the cyclic code $\mathcal{C}$ is determined in Theorem \ref{W02}.

\begin{lemma} \label{SM03}
Let $p=3, q=p^m$ where $m\geq 6$ is an even integer. The notations defined in equations (\ref{0001}) and (\ref{0002}) satisfy the following equations
\[
\begin{array}{lll}
n_{1,0}&=&-\{b + c_6 - ap^2 - ap^3 - ap^4 - ap^5 - bp^2 + c_1p^6 \\
    &+& c_2p^6 + c_3p^3 + c_3p^5 - c_4p^2 \\
    &-& c_4p^4 + c_5p^2\}/{(- 2p^6 + 2p^4 + 2p^2 - 2)} \\
n_{-1,0}&=&-\{b - c_6 - ap^2 - ap^3 - ap^4 - ap^5 - bp^2 + c_1p^6\\
& -& c_2p^6 + c_3p^3 + c_3p^5\\
 &+& c_4p^2 + c_4p^4 + c_5p^2\}/{(- 2p^6 + 2p^4 + 2p^2 - 2)} \\
2n_1&=&{{-(b - c_5 + ap^3 - c_3p^3)}\over {(p^2 - p^4)}}\\
n_{1,2}&=&\{c_4 - c_6 + ap - c_5p + ap^2 + ap^4 + ap^5 - c_1p^5\\
& - &c_2p^4 - c_3p^2 - c_3p^4 + c_4p^4\}/{(2p^7 - 4p^5 + 2p^3)}\\
n_{-1,2}&=&-\{c_4 - c_6 - ap + c_5p - ap^2 - ap^4 - ap^5 + c_1p^5 \\
&-& c_2p^4 + c_3p^2 + c_3p^4 + c_4p^4\}/ {(2p^7 - 4p^5 + 2p^3)}\\
2n_3&=&{{(b - c_5 + ap - c_3p)}\over {(p^4 - p^6)}}\\
n_{1,4}&=&-\{b + c_4 - c_5 - c_6 + ap - c_3p + ap^2 + ap^3\\
& +& ap^4 - bp^2 - c_1p^4 \\
&-& c_2p^2 - c_3p^3 + c_4p^2\}/{ (2p^{10} - 2p^8 - 2p^6 + 2p^4)}\\
 n_{-1,4}&=& -\{b - c_4 - c_5 + c_6 + ap - c_3p + ap^2 + ap^3 \\
 &+& ap^4 - bp^2 - c_1p^4 \\
 &+& c_2p^2 - c_3p^3 - c_4p^2\}/{(2p^{10} - 2p^8 - 2p^6 + 2p^4)}
\end{array}
\]
where
\[
\begin{array}{ll}
c_1=p^{3m}-1, c_2=p^{{m}\over 2}\left(p^{2m}-1\right), c_3=p^{m} \left(p^m-1\right), \\
c_4= (p+1)p^{{3m}\over 2}\left(p^{m}-1\right), c_5=\left(8(p^m-1)^2-p^m+1\right)p^m, \\
 c_6=\left(5(p^m-1)(8p^m-2p-10)-p^{2m}+1\right)p^{m\over 2},
\end{array}
\]
and $a, b$ are defined in Lemma \ref{SM02} which needs $m\geq 6$.
\end{lemma}

\begin{IEEEproof}
From Lemma \ref{ES002}, Corollary \ref{C02}, Corollary \ref{C03} and Lemma \ref{SM02}, the notations satisfy the following equations
\[
\left\{
\begin{array}{lll}
&&2(n_1+n_3)+n_{-1,0}+n_{1,0}+n_{-1,2}+n_{1,2}\\
&&+n_{1,4}+n_{-1,4} \\
& =&c_1,\\
  &&n_{1,0}-n_{-1,0}+p(n_{1,2}-n_{-1,2}) +p^2(n_{1,4}-n_{-1,4})\\
    &=&c_2, \\
&& -2\left(pn_1+p^3n_3\right)+n_{1,0}+n_{-1,0}+p^2(n_{1,2}+n_{-1,2})\\
&&+p^{4}(n_{1,4}+n_{-1,4}) \\
&=& c_3,\\
  && n_{1,0}-n_{-1,0}+p^3(n_{1,2}-n_{-1,2})+p^6(n_{1,4}-n_{-1,4}) \\
   &=&c_4,\\
  && n_{1,0}+n_{-1,0}+2n_1p^2+p^4(n_{1,2}+n_{-1,2})+2n_3p^6\\
  &&+p^8(n_{1,4}+n_{-1,4})\\
  &=&c_5,\\
  && n_{1,0}-n_{-1,0}+p^5(n_{1,2}-n_{-1,2})+ p^{10}(n_{1,4}-n_{-1,4})\\
  &=&c_6,\\
   &&n_{1,0}+n_{-1,0}+p^2(n_{1,2}+n_{-1,2})+p^4(n_{1,4}+n_{-1,4})\\
   &=&a,\\
 &&  n_{1,0}+n_{-1,0}+p^4(n_{1,2}+n_{-1,2})+p^8(n_{1,4}+n_{-1,4})\\
 &=&b.
\end{array}
\right.
\]
Solving the above equation system, the result is obtained.
\end{IEEEproof}


\begin{lemma}
Let $p=3, q=p^m$ where $m\geq 6$ is an even integer. 
The number of solutions of the equation system
\begin{equation}\label{M05}
\left\{
\begin{array}{ll}
x^{2}+y^2+z^2+w^2+u^2+v^2 &=0\\
x^{p+1}+y^{p+1}+z^{p+1}+w^{p+1}+u^{p+1}+v^{p+1} &=0\\
x^{p^2+1}+y^{p^2+1}+z^{p^2+1}+w^{p^2+1}+u^{p^2+1}+v^{p^2+1} &=0,
\end{array}
\right.
\end{equation}
is
\[
\begin{array}{ll}
&n_{1,0}+n_{-1,0}+p^6(n_{1,2}+n_{-1,2})+p^{12}(n_{1,4}+n_{-1,4})\\
&-p^3(n_{1,1}+n_{-1,1})-p^9(n_{1,3}+n_{-1,3})+p^{3m}.
\end{array}
\]
\end{lemma}

\begin{IEEEproof}
The following moment of exponential sum $S(\alpha,\beta,\gamma)$ satisfies
\[
\begin{array}{lll}
&&\sum\limits_{\alpha,\beta,\gamma \in \mathbb{F}_q}S(\alpha,\beta,\gamma)^6\\
&=& p^{3m}(n_{1,0}+n_{-1,0})+p^{3(m+2)}(n_{1,2}+n_{-1,2})\\
&+&p^{3(m+4)}(n_{1,4}+n_{-1,4})-p^{3(m+1)}(n_{1,1}+n_{-1,1})\\
&-&p^{3(m+3)}(n_{1,3}+n_{-1,3})+p^{6m}\\
&=&M_6\cdot p^{3m}
\end{array}
\]
where $M_6$ is the number of solutions of (\ref{M05}). Solving the above equation for $M_6$, the result is obtained.
\end{IEEEproof}
Equation (\ref{M05}) considers the case for $6$ variables. Using the seventh moment of $S(\alpha,\beta,\gamma)$, the number of solutions can be calculated when there are $7$ variables, etc.

\begin{theorem}\label{W02}
Let $p=3, q=p^m$ where $m\geq 6$ is an even integer. The cyclic code $\mathcal{C}$ with nonzeros $\pi^{-2}, \pi^{-(p+1)}$ and $\pi^{-(p^2+1)}$
has seven nonzero weights,
\begin{equation}
\begin{array}{lll}
&A_{p^{m-1}(p-1)}=2(n_1+n_3), A_{p^{m-1}(p-1)-{{p-1}\over p}{p^{m\over 2}}}=n_{1,0}, \\
&A_{p^{m-1}(p-1)+{{p-1}\over p}{p^{m\over 2}}}=n_{-1,0},A_{p^{m-1}(p-1)-{{p-1}\over p}{p^{{m+2}\over 2}}}=n_{1,2},\\
 &A_{p^{m-1}(p-1)+{{p-1}\over p}{p^{{m+2}\over 2}}}=n_{-1,2},\\
 & A_{p^{m-1}(p-1)-{{p-1}\over p}{p^{{m+4}\over 2}}}=n_{1,4}, \\
 & A_{p^{m-1}(p-1)+{{p-1}\over p}{p^{{m+4}\over 2}}}=n_{-1,4}.
\end{array}
\end{equation}
where $\pi$ is a primitive element of the finite field $\mathbb{F}_q$.
\end{theorem}

It is interesting to note about the weights of the cyclic code $\mathcal{C}$. If there is a weight of the form $p^{m-1}(p-1)+{{p-1}\over p}{p^{{m+2i}\over 2}} (i=0,1,2)$, then there is a weight of the form $p^{m-1}(p-1)-{{p-1}\over p}{p^{{m+2i}\over 2}}$. So, it seemed as if the weights are symmetric about the value $p^{m-1}(p-1)$ which is also a weight of $\mathcal{C}$. As the following example illustrates, in general the higher the value $i$ the less number of corresponding weights. This phenomenon may be explained to the fact that the linear part in the exponential position of the parameter $\zeta$ acts as a center role in the formation of the weights.

\begin{example}
Let $p=3, q=p^m$ where $m=6$. The cyclic code $\mathcal{C}$ has nonzeros $\pi^{-2}, \pi^{-(p+1)}$ and $\pi^{-(p^2+1)}$ where $\pi$ is a primitive element of the finite field $\mathbb{F}_q$. Using Matlab, it can be found that it has seven nonzero weights
\[
\begin{array}{ll}
&A_{486}=124245576, A_{468}=128432304,\\
& A_{504}=119277522, A_{432}=8591310, \\
& A_{540}=6866496, A_{324}=4732, A_{648}=2548,
\end{array}
\]
which is verified by using Matlab.
\end{example}

\section{Conclusion}\label{SecIV}

Since the weight distributions played an important role in the application of cyclic codes, this paper focuses on the determination of a class of cyclic codes with three nonzeros. Relevant results received a lot of attention by using methods with lower moments of exponential sum.
Here we try to apply higher moments to deal with the problem.




\ifCLASSOPTIONcaptionsoff
  \newpage
\fi

%





\begin{thebibliography}{1}


\bibitem{BC001}
 W. Bosma, J. Cannon, and C. Playoust, ``The Magma algebra system. I. The user language,'' {\sl J. Symbolic Comput.}, vol. 24, no. 3-4,
pp. 235--265, 1997.

\bibitem{BM001}
N. Boston and G. McGuire, ``The weight distributions of cyclic codes with two zeros and zeta
functions,'' {\sl J. Symbolic Comput.}, vol. 45, no. 7, pp. 723--733, 2010.


 \bibitem{CH001}
Y. Chen and A. J. Han Vinck, ``A lower bound on the optimum distance profiles of the second-order Reed-Muller codes,''
   {\sl IEEE Trans. Inf. Theory}, vol. 56, no. 9, pp. 4309--4320, Sep. 2010.

\bibitem{D001}
  C. Ding, ``The weight distribution of some irreducible cyclic codes,''
{\sl IEEE Trans. Inf. Theory}, vol. {55}, no. 3, pp. 955--960, Mar. 2009.

\bibitem{DL001}
 C. Ding, Y. Liu, C. Ma, and L. Zeng, ``The weight distributions of the duals of cyclic codes with two zeros,'' {\sl IEEE Trans. Inf.
Theory}, vol. 57, no. 12, pp. 8000--8006, Dec. 2011.

\bibitem{DY001}
 C. Ding and J. Yang, ``Hamming weights in irreducible cyclic codes,'' {Discrete Mathematics}, vol. 313, no. 4, pp. 434--446, 2013.

\bibitem{FL001}
 K. Feng and J. Luo, ``Weight distribution of some reducible cyclic codes,'' {\sl Finite Fields Appl.}, vol. 14, no. 4, pp. 390--409, Apri. 2008.

\bibitem{FY001}
R. W. Fitzgerald and J. L. Yucas, ``Sums of Gauss sums and weights of
irreducible codes,''
{\sl Finite Fields Appl.}, vol. {11}, no. 1, pp. 89--110, Jan. 2005.

\bibitem{JH001}
 A. Johansen and T. Helleseth, ``A family of $m$-sequences with five valued
cross correlation,''
{\sl IEEE Trans. Inf. Theory}, vol. 55, no. 2, pp. 880--887, Feb. 2009. 

\bibitem{JHK001}
 A. Johansen, T. Helleseth, and A. Kholosha, ``Further results on m-sequences with five-valued cross correlation,''
 {\sl IEEE Trans. Inf. Theory,} vol. 55, no. 12, pp. 5792--5802, Dec. 2009. 

\bibitem{K001}
 T. Kasami, ``A decoding procedure for multiple-error-correcting cyclic
codes,'' {\sl IEEE Trans. Inf. Theory}, vol. 10, no. 2, pp. 134--138, Apri. 1964.





\bibitem{LHF001}
 S. X. Li, S. H. Hu, T. Feng, and G. Ge, ``The weight distribution of a class of cyclic codes related to Hermitian forms Graphs,''
arXiv:1212.6371, 2012.

\bibitem{LH001}
  R. Lidl and H. Niederreiter, {\sl Finite Fields},
 Cambridge, U.K.: Cambridge
Univ. Press, 1997.

\bibitem{LL001}
 X. Liu and Y.  Luo, ``On the bounds and achievability about the ODPC of $\mathcal{GRM}(2,m)^*$ over prime field
for increasing message length,''
in manuscript.

\bibitem{LL002}
X. Liu and Y. Luo, ``The weight distributions of some cyclic codes with three or four nonzeros over $\mathbb{F}_3$,''
arXiv:1302.0394v1. 



\bibitem{LTW001}
 J. Luo, Y. Tang, and H. Wang, ``Cyclic codes and sequences: the generalized Kasami case,'' {\sl IEEE Trans. Inf. Theory}, vol. 56, no.
12, pp. 2130--2142, May. 2010.


\bibitem{MS01}
 F. J. MacWilliams and J. Seery, ``The weight distributions of some minimal cyclic codes,''
 {\sl IEEE Trans. Inf. Theory}, vol. {27}, no. 6, pp. 796--806, 1981.

\bibitem{MS001}
F. J. MacWilliams and N. J. A. Sloane,
{\sl The Theory of Error-Correcting Codes.}
North-Holland, Amsterdam, 1988.


 \bibitem{M001}
  R. J. McEliece, ``Irreducible Cyclic Codes and Gauss
Sums,'' in:{\sl Combinatorics, Part I: Theory of Designs, Finite Geometry
and Coding Theory}.
  Math. Centre Tracts, Amsterdam: Math. Centrum, no. {55}, pp. 179--196, 1974.


\bibitem{MR001}
R. J. McEliece and H. Rumsey, Jr., ``Euler products, cyclotomy, and
coding,'' {\sl J. Number Theory,} vol. 4, pp. 302--311, 1972.


\bibitem{PH001}
V. Pless and W.C. Huffman, {\sl Handbook of Coding Theory.} Elsevier, Amsterdam, 1998.





 \bibitem{S001}
 I. R. Shafarevich, Basic Algebraic Geometry 1. Springer, 2nd ed., 1994.


\bibitem{SF001}
S. G. S. Shiva and K. C. Fung, ``Permutation decoding on certain tripleerror-
correcting binary codes,'' {\sl IEEE Trans. Inf. Theory}, vol. 18, no. 3,
pp. 444--446, 1972.

\bibitem{SFT001}
S. G. S. Shiva, K. C. Fung, and H. S. Y. Tan, ``On permutation decoding
of binary cyclic double-error-correcting codes of certain lengths (corresp.),''
{\sl IEEE Trans. Inf. Theory}, vol. 16, no. 5, pp. 641--643, 1970.

\bibitem{V01}
M. Van Der Vlugt, ``Hasse-Davenport curve, Gauss sums and weight
distribution of irreducible cyclic codes,'' {\sl J. Number Theory,} vol. 55, pp.
145--159, 1995.


\bibitem{V001}
M. Van Der Vlugt, ``Surfaces and the weight distribution of a family of
codes,'' {\sl IEEE Trans. Inf. Theory}, vol. 43, no. 4, pp. 1354--1360, Apr. 1997.


\bibitem{X002}
 M. Xiong, ``The weight distributions of a class of cyclic codes II,'' to appear in Des. Codes Cryptogr.,





\bibitem{ZDJZ001}
Z. Zhou, C. Ding, J. Luo, and A. Zhang, ``A Family of Five-Weight Cyclic
Codes and Their Weight Enumerators,'' arXiv:1302.0952v1.









\end{thebibliography}
\end{document}